\documentstyle[12pt]{article}
\begin{document}
\def\lsim{\:\raisebox{-0.5ex}{$\stackrel{\textstyle<}{\sim}$}\:}
\def\gsim{\:\raisebox{-0.5ex}{$\stackrel{\textstyle>}{\sim}$}\:}
\def\hw{\hbar \omega}
\def\d{\dagger}
\def\ad{ {\bf a}^{\dagger}}
\def\ba#1{\begin{array}{#1}}
\def\ea{\end{array}}
\def\bin#1#2{\frac{#1!}{(#1-#2)! #2!}}
\def\hre#1#2{\frac{#1!}{(#1-2#2)! #2!}}
\def\hreg#1#2#3{\frac{#1!}{#2! (#1-#3 #2)!}}
\def\nsp{\negthinspace}
\def\b#1#2{\left(\nsp \begin{array}{r}#1\\#2\end{array} \nsp \right)}
\def\hr#1#2{\left[\nsp \begin{array}{c} #1\\ #2 \end{array} \nsp \right]}
\def\hrg#1#2#3{\left[\nsp\ba{c}#1\\#2\ea \nsp \right]_{#3}}
\def\above#1#2{{\stackrel {{\textstyle #1}}{\scriptstyle { #2}}}}
\def\pb{\beta^{'} }
\def\be{\begin{equation}}
\def\ee{\end{equation}}
\def\br{\begin{eqnarray}}
\def\er{\end{eqnarray}}
\def\x{\times}
\def\hs{{\cal H}}
\def\go{\rightarrow  }
\def\abs#1{{\mid #1 \mid }}
\def\la{\langle }
\def\kg{\preceq }
\def\ra{\rangle }
\def\lb#1{{\label{eq:#1} }}
\def\rf#1{{(\ref{eq:#1}) }}
\def\dox{$\times$ }
\def\dodt{{$\bullet $ }}
\def\bra#1{\langle #1|}
\def\ket#1{|#1 \rangle}
\def\bra#1{\langle #1|}
\def\ex#1#2{\langle #1 | #2 | #1 \rangle }
\def\sp#1{\langle #1 \rangle }
\def\ov#1#2{\langle #1 | #2  \rangle }
\def\exp#1#2#3{\langle #1 | #2 | #3 \rangle }
\def\E {{E^+}}
\def\A {{{\cal A}}}
\def\F {{{\cal F}}}
\def\G {{{\cal G}}}
\def\I {{{\cal I}}}
\def\K {{{\cal K}}}
\def\M {{{\cal M}}}
\def\N {{{\cal N}}}
\def\O {{{\cal O}}}
\def\Q {{{\cal Q}}}
\def\R {{{\cal R}}}
\def\S {{{\cal S}}}
\def\X {{{\cal X}}}
\def\Bra#1{\langle #1||}
\def\Ket#1{||#1 \rangle}
\def\beg#1#2{\left(\nsp \begin{array}{r}#1\\#2\end{array} \nsp \right)}
\begin{titlepage}
\pagestyle{empty}
\baselineskip=21pt
\vskip .2in
\begin{center}
{\large{\bf On the energy-shell contributions of the three-particle~-~
three-hole excitations }} \end{center}
\vskip .1in
\begin{center}

A. Mariano$^{\d}$ and F. Krmpoti\'{c}$^{\d}$

{\small\it Departamento de F\'\i sica, Facultad de Ciencias Exactas,}\\
{\small\it Universidad Nacional de La Plata, C. C. 67, 1900 La Plata,
Argentina}

{\small\it and}

A.F.R. de Toledo Piza

{\small\it Instituto de F\'\i sica,
Universidade de S\~{a}o Paulo
C.P. 20516, 01498 S\~{a}o Paulo, Brasil}

\end{center}
\vskip 0.5in
\centerline{ {\bf Abstract} }
\baselineskip=18pt
The response functions for the extended second and third random
phase approximation are compared. A second order perturbation
calculation shows that the first-order amplitude for the direct
$3p3h$ excitation from the ground state cancels with those that
are engendered by the  $1p1h$-$3p3h$ coupling. As a consequence
nonvanishing $3p3h$ effects to the $1p1h$ response involve off
energy shell renormalization only. On shell $3p3h$ processes are
absent.

\vspace{1.5in}

\noindent
$^{\dagger}$Fellow of the CONICET from Argentina.
\end{titlepage}
\baselineskip=18pt

Many efforts have been devoted during the last few years in developing
generalized random phase approximations (RPA), which go beyond the standard
one-particle - one-hole ($1p1h$) approach \cite{Row}. This has been
accomplished by including additional correlation effects in both the ground
state and the excited states
\cite{Ber,Yan,Alb1,Tak1,Dro1,Mac,Ada,Nis,Tak2,Dro2,Alb2,Mar1,Tak3,Mar2}.
The reasons for that were mainly: i) the problem of the missing strength in
the Gamow-Teller (GT) resonances, induced by $(p,n)$ reactions
\cite{Gar,Rap},
and ii) the issue of the missing charge and missing dip-strength in
quasielastic electron scattering \cite{Bar}.
In particular, the extended second RPA (ESRPA), which explicitly includes
the $2p2h$ ground state correlations (GSC), was extensively used to describe
the above mentioned nuclear excitations
\cite{Ber,Alb1,Tak1,Dro1,Nis,Tak2,Dro2,Alb2,Tak3}.
Yet, it is self evident that when the $2p2h$ admixtures  are present in the
ground state, the external excitation field can lead, not only to the $1p1h$
and $2p2h$ states in the final nucleus, but also to the $3p3h$ states.
However, as the ESRPA does not involve the $3p3h$ propagator these excitations
cannot appear within the response function
as  real on the energy-shell processes.
Recently the $3p3h$ degrees of freedom were explicitly included within a
Tamm-Dancoff approach (TDA), and their effects on the non-energy-weighted GT
sum-rule were discussed \cite{Mar1}.
Also an extended third RPA (ETRPA), which possesses as the TDA limit
the formalism developed in ref. \cite{Mar1}, has been used to study
the effects of $3p3h$ excitations on the static strength function
for quasielastic electron scattering \cite{Mar2}.

The purpose of this paper is to present some results for on the energy-shell
$3p3h$ effects in the response function. This is done in the context of the
full ETRPA approach which is therefore reviewed below. The nature of the
resulting response function is then confronted to what one obtains using
the ESRPA by performing a perturbative expansion of the responses in each
case. The possibility of having a three nucleon ejection process is finally
analyzed in this framework.

Let us start with the linear response to an external field $\hat{F}$
defined as
\be
R(E) = -i \int_{-\infty}^{\infty} \bra{\tilde{0}}T\left[\hat{F}^{H\dag}(t)
\hat{F}^{H}(0)\right]\ket{\tilde{0}} e^{iEt} dt,
\label{1}
\ee
where $\hat{F}^H(t)\equiv e^{i\hat{H}t}\hat{F}e^{-i\hat{H}t}$,
$\hat{H}=\hat{H}_{0}+\hat{V}$, with $\hat{H}_{0}$ and $\hat{V}$ being,
respectively, the Hartree-Fock (HF) mean
field and  the residual interaction.
The spectral representation of the response function, in terms of a set
$\{\ket{\nu}\}$ of eigenstates of the hamiltonian $\hat{H}$, reads
\be
R(E)=\sum_{\nu}\left[
 \frac{\bra{\tilde{0}}\hat{F}\ket{\nu}\bra{\nu}\hat{F}^{\d}
\ket{\tilde{0}}}{E-E_{\nu}+i\eta}
-\frac{\bra{\tilde{0}}\hat{F}^{\d}\ket{\nu}\bra{\nu}\hat{F}
\ket{\tilde{0}}}{E+E_{\nu}-i\eta}
\right],
\label{2}
\ee
where $\eta$ is an infinitesimal positive number.

Within the equation of motion method \cite{Row},
the set $\{\ket{\nu}\}$ is generated as
\be
\ket{\nu}= \Omega_{\nu}^{\d}\ket{\tilde{0}};
\hspace{2mm}\Omega_{\nu}^{\d} = \sum_{i} X_{i}^{\nu} C_{i}^{\d}
- \sum_{j} Y_{j} ^{\nu}C_{j},
\label{3}
\ee
and
\be
\Omega_{\nu} \ket{\tilde{0}}= 0, \hspace{10mm} \mbox{for all $\nu$}.
\label{4}
\ee
The operators $C_{i}^{\d}$ and $C_{i}$
(with $C_i^{\d}\equiv a_{p_1}^{\d}\cdot \cdot \cdot a_{p_i}^{\d}a_{h_1}
\cdot \cdot \cdot a_{h_i}$) create and annihilate $i$
particle-hole pairs on the HF vacuum $\ket{0}\equiv \ket{0p0h}$, respectively.

The equation of motion for $\Omega_\nu^{\d}$

\be
\bra{\tilde{0}}\left[\Omega_{\nu},\left[H,\Omega_{\mu}^{\d}\right]\right]\ket{\tilde{0}}
=
E_{\nu}\bra{\tilde{0}}\left[\Omega_{\nu},\Omega_{\mu}^{\d}\right]\ket{\tilde{0}}
\delta_{\nu,\mu},
\label{5}
\ee
where $E_{\nu}$ stands for the excitation energy of the state $\ket{\nu}$,
leads to the RPA-like eigenvalue problem
\be
\A \X^\nu = E_{\nu} \N \X^{\nu},
\label{6}
\ee
with
\be
\A = \left(\begin{array}{cc} A & B \\ B^{\ast} & A^{\ast}\end{array}\right),
\hspace{5mm}
\X^{\nu} = \left(\begin{array}{l} X^{\nu}\\Y^{\nu}\end{array}\right),
\hspace{5mm}
\N = \left(\begin{array}{cc} N & 0\\ 0 &-N^{\ast}\end{array}\right).
\label{7}
\ee
The submatrices $A$, $B$ and $N$ given by
\be
A_{i,j} =
\bra{\tilde{0}}\left[C_{i},\left[H,C_{j}^{\d}\right]\right]\ket{\tilde{0}},
\hspace{2mm}
B_{i,j} =
\bra{\tilde{0}}\left[C_{i},\left[H,C_{j}\right]\right]\ket{\tilde{0}},
\hspace{2mm}
N_{i,j} = \bra{\tilde{0}}\left[C_{i},C_{j}^{\d}\right]\ket{\tilde{0}},
\label{8}
\ee
and using eqs. (\ref{2}-\ref{6}) it is possible to write the response in
representation independent form as
\be
R(E) = \F^{\d} ( E \N - \A+i\eta \I)^{-1} \F,
\label{9}
\ee
where  $\F$ is defined as
\be
\F\equiv \left(\begin{array} {ll} F^{A} \\ F^{B} \end{array}\right),
\mbox{with}
\left\{ \begin{array}{l}
F^{A}_{i}=\bra{\tilde{0}}\left[ C_{i},\hat{F}\right]\ket{\tilde{0}},\\
F^{B}_{i}
= F_{i}^{A\ast}(\hat{F} \go \hat{F}^{\d}).
\end{array}\right.
\label{10}
\ee

After splitting the Hilbert space of $ipih$ states into a P-space that
includes only the $1p1h$ states and the Q-space that spans on the rest of
the states, the response function can be written as
\be
R(E) = \tilde{\F}_P^{\d}(E)\G_P(E) \tilde{\F}_P(E)+ \F_Q^{\d}\G_Q(E)\F_Q,
\label{11}
\ee
where
\be
\G_P(E)=\left[E \N_P+i\eta \I_P-\A_P -\left(\A_{PQ}-\N_{PQ}E \right)
\G_Q(E) \left(\A_{QP}-\N_{QP}E \right)\right]^{-1},
\label{12}
\ee
with
\be
\G_Q(E)=\left[E\N_Q+i\eta \I_Q-\A_Q \right]^{-1},
\label{13}
\ee
and
\be
\tilde{\F}_{P}(E)=\F_P-\N_{PQ}\F_Q+\A_{PQ}\G_{Q}(E)\F_Q.
\label{14}
\ee

In  standard RPA the state $\ket{\tilde{0}}$ is approximated by the HF ground
state and the Q-space is absent, while the so called extended RPA incorporates
perturbative ground state $2p2h$  admixtures and a perturbatively suggested
truncation of the dynamical matrices and excitation operator. It is
obtained by:\\
i) evaluating the matrix elements (\ref{8}) and (\ref{10}) for \cite{Tak2}
\be
\ket{\tilde{0}} = c_0\ket{0} + \sum_{2_0} c_{2_0}\ket{2_0},
\label{15}
\ee
where
\be
c_0 \cong 1-\frac{1}{2}\sum_{2_0}\left|c_{2_0}\right|^2,
\hspace{2mm}
c_{2_0}\cong-\frac{V_{2_00}}{E_{2_0}},
\label{16}
\ee
$2_{0}\equiv (p_1p_2h_1h_2)_0$ represents the $2p2h$ ground state admixtures,
$E_{2_0}$ the corresponding unperturbed energy and
$V_{2_00}\equiv \bra{2_0}V\ket{0}$, and\\
ii) keeping terms up to second order in $\hat{V}$ for the forward sector
within the P space, terms linear in $\hat{V}$ for the backward sector within
the P space and for the coupling between the P and Q spaces, and only
terms of zeroth order within the Q space.
Under these conditions the norm matrix elements read \cite{Tak1})
\be
N_{ij}=\delta_{ij} + \Delta N_{ij}
\label{17}
\ee
where $i\equiv ipih$ and the nonzero $\Delta N_{ij}$ are
\be
\Delta N_{11'}=
\sum_{2_0,2_0'}c_{2_0}^{\ast}c_{2_{0}'} \bra {2_0}\hat{D}_{11'}\ket{ 2_{0}'},
\hspace{2mm}
\Delta N_{13}  = \sum_{2_0}c_{2_0}^{\ast} \ov {1;2_0}{3},
\label{18}
\ee
where $\hat{D}_{11'}=\left[\hat{C}_1,\hat{C}_{1'}^{\d}\right]-\delta_{11'}$
and $\ov {1;2_0}{3}$ is the overlap between the $1p1h\otimes(2p2h)_0$ and
$3p3h$
final state configurations.
(Note that within the quasi-boson approximation $\hat{D}_{11'}\equiv 0$.)
The explicit result for the matrix element
$\bra{2_0}\hat{D}_{11'} \ket{2_0'}$ is
\br
&&\bra{(p_1p_2h_1h_2)_0}\hat{D}_{ph,p'h'}\ket{(p'_1p'_2h'_1h'_2)_0}
=-\left[1+P(h_1,h_2)P(h'_{1},h'_{2})\right]\\
\nonumber
&\x& \left[\delta_{p,p'}\delta_{h_1,h'} P^-(h,h_2)P^-(p_1,p_2)
\delta_{h'_{1},h}\delta_{h_2,h'_{2}}\delta_{p_2,p'_{2}}\delta_{p_1,p'_{1}}
\right] + p \leftrightarrow h ,
\label{19}
\er
where $P^-(i,j)\equiv [ 1 - P(i,j)]$, while the operator $P(i,j)$
exchanges the arguments $i$ and $j$.

The forward going energy matrix elements are evaluated in the same way
and one gets
\be
A_{ij}=\delta_{ij}E_j + V_{ij} +\Delta A_{ij},
\label{20}
\ee
where $V_{ij}\equiv \bra{i}\hat{V}\ket{j}$ and
the nonzero matrix elements $\Delta A_{ij}$ are:
\be
\Delta A_{11'}=\sum_{2_0,2_{0}'}(E_1-E_{2_0})c_{2_0}^{\ast}c_{2_{0}'}
\bra{2_0}\hat{D}_{11'}\ket{2_{0}'},
\hspace{2mm}
\Delta A_{13}=\Delta N_{13}E_3.
\label{21}
\ee

The one-body matrix elements are:
\be
F_i^A =  \left\{ \begin{array}{l}
f_1 + \sum_{1'} \Delta N_{11'}f_{1'} \hspace{5mm} \mbox{for $i=1$}
\vspace{5mm}
\\
\sum_{2_0} c_{2_0} f_{i2_0}\hspace{11mm}\mbox{for $i>1$,}
\end{array}\right.
\label{22}
\ee
where
\be
f_1 \equiv \bra{1}\hat{F}\ket{0}, \hspace{2mm} \mbox{and} \hspace{2mm}
f_{i2_0}\equiv \bra{i}\hat{F}\ket{2_0}.
\label{23}
\ee

Before proceeding it is convenient to introduce the unperturbed Green's
function:
\be
\G^0(E)\equiv  \left(\begin{array}{cc} G^0(E) & 0 \\
0 & G^{0\ast}(-E) \end{array}
\right),
\label{24}
\ee
where $G^0(E) \equiv [ \E - A(\hat{H}=\hat{H_0}) ]^{-1}$
(with $\E\equiv E+i\eta$) and
rewrite the perturbed Green function within the space $P$ in the form
\be
\G_P(E)=\left[\left(\G^0_P(E)\right)^{-1} - \K_P(E)\right] ^{-1},
\label{25}
\ee
where
\be
\K_P(E)\equiv \K_{\sf 11'}(E) =
\left(\begin{array}{cc} V_{\sf 11'} + \Sigma_{\sf 11'}(E)& B_{\sf 11'}\\
B^{\ast}_{\sf 11'} & V_{\sf 11'}^{\ast} + \Sigma_{\sf 11'}^{\ast}(-E)
\end{array} \right),
\label{26}
\ee
with
\be
\Sigma_{\sf 11'}(E)=\Delta\Sigma_{\sf 11'}^{(2)}(E) +
\Delta\Sigma_{\sf 11'}^{(3)}(E)+
\sum_{{\sf i}={\sf 2,3}} V_{\sf 1i}G^0_{\sf ii}(E)V_{\sf i1'},
\label{27}
\ee
and
\br
\Delta\Sigma_{\sf 11'}^{(2)}(E)&=& \Delta A_{\sf 11'} - \Delta N_{\sf 11'} E ,
\nonumber \\
\Delta\Sigma_{\sf 11'}^{(3)}(E)&=& -
\left(2V_{\sf 13}-\Delta N_{\sf 13}\left(G^0_{\sf 33}(E)\right)^{-1}\right)
\Delta N_{\sf 31'}.
\label{28}
\er
In the above equations $V_{\sf ij'}$ stands for the matrix representation
of the residual interaction within the $ipih \otimes jpjh$ subspace.

The response function now reads
\be
R(E)=\tilde{\F}_{\sf 1'}(E)
\G_{\sf 11'}(E)\tilde{\F}_{\sf 1'}(E)
+\sum_{\sf i=2,3} {\F}_{\sf i}^{\d}\G^0_{\sf ii}(E){\F}_{\sf i},
\label{29}
\ee
where
\be
\tilde{\F}_{\sf 1}(E)\equiv \left(\begin{array}{ll} \tilde{F}^{A}_{\sf 1}(E)
 \\ \tilde{F}^{B}_{\sf 1}(E) \end{array}\right),
\mbox{with}
\left\{ \begin{array}{l}
\tilde{F}^{A}_{\sf 1}(E)=f_{\sf 1}+
\Delta\tilde{F}_{\sf 1}(E),
\vspace{5mm}
\\
\Delta\tilde{F}_{\sf 1}(E)=
\Delta F^{(2)}_{\sf 1}+ \Delta F^{(3)}_{\sf 1} +
\sum_{{\sf i}={\sf 2,3}} V_{\sf 1i}G^0_{\sf ii}(E) F_{\sf i}
\vspace{5mm}
\\
\Delta F^{(2)}_{\sf 1} = \Delta N_{\sf 11'} f_{\sf 1'},~~~
\Delta F^{(3)}_{\sf 1} = - \Delta N_{\sf 13} F_{\sf 3}.
\end{array}\right.
\label{30}
\ee

{}From the expressions for $\Delta A_{11'}$ and $\Delta N_{11'}$,
given by Eqs. (\ref{18}) and (\ref {21}) respectively, the matrix elements
$\Delta\Sigma_{\sf 11'}^{(2)}(E)$ and $\Delta \tilde{F}_1^{(2)}$ can be
expressed as:
\br
\Delta \Sigma_{11'}^{(2)}(E)& =&-\sum_{2_0,2_0'}c_{2_0}^{\ast}c_{2_0'}
\bra{2_0}\hat{D}_{11'}\ket{2_0'}(E - E_1 + E_{2_0'}),
\label{31}\\
\Delta \tilde{F}_1^{(2)}&=&\sum_{2_0,2_0'}c_{2_0}^{\ast}c_{2_{0}'}
\bra {2_0}\hat{D}_{11'}\ket{ 2_{0}'}f_{1'} .
\label{32}
\er
Moreover, from the relationships
\be
V_{13} =  - \sum_{2_0} c_{2_0}^* E_{2_0} \ov{1; 2_0}{3}; \hspace{3mm}
f_{3,2_{0}}= \sum_{1} \ov{3}{1; 2_{0}} f_1 ,
\label{33}
\ee
one obtains
\br
\Delta \Sigma_{11'}^{(3)}
& = & \sum_{2_0,2_0'}c_{2_0}^{\ast}c_{2_0'}
\ov{1;2_0}{1';2_0'}(E - E_1 + E_{2_0'}),
\label{34}\\
\Delta \tilde{F}_1^{(3)}
&=&-\sum_{2_0,2_0'}c_{2_0}^{\ast}c_{2_{0}'}
\ov {1;2_0}{1';2_{0}'}f_{1'}.
\label{35}
\er
We can note here that
\br
\ov {1;2_0}{1';2_{0}'}
=\bra {2_0}(\hat{D}_{11'}+
\hat{d}_{11'})\ket{ 2_{0}'},\hspace{2mm}\mbox{with}\hspace{2mm}
\hat{d}_{11'}\equiv
\delta_{11'}+C^{\d}_{1'}C_1,
\label{36}
\er
and thus in summary we get:

i) in the ESRPA (where the $Q$ space
includes only the $2p2h$ excitations)
\br
\Sigma_{11'}(E)&=&-\sum_{2_0,2_0'}c_{2_0}^{\ast}c_{2_0'}
\bra{2_0}\hat{D}_{11'}\ket{2_0'}(E - E_1 + E_{2_0'})
+\sum_2 \frac{V_{12} V_{21'}}{\E -E_2},
\label {37}\\
\tilde{\F}_1(E)&=& f_1 + \sum_{2_0,2_0';1'}c_{2_0}^{\ast}c_{2_{0}'}
\bra {2_0}\hat{D}_{11'}\ket{ 2_{0}'}f_{1'} +
\sum_{2,2_0;} \frac{V_{12} f_{22_0} c_{2_0}}{\E -E_2}
\label{38};
\er
ii) in the ETRPA (where the $Q$ space
includes both the $2p2h$ and $3p3h$ excitations)
\br
\Sigma_{11'}(E)&=&\sum_{2_0,2_0'}c_{2_0}^{\ast}c_{2_0'}
\bra{2_0}\hat{d}_{11'}\ket{2_0'}(E - E_1 + E_{2_0'})
+\sum_{i= 2,3} \frac{V_{ 1i}V_{i1'}}{\E -E_i},
\label{39} \\
\tilde{\F}_1(E)&=& f_1
- \sum_{2_0,2_0'}c_{2_0}^{\ast}c_{2_{0}'}
\bra {2_0}\hat{d}_{11'}\ket{ 2_{0}'}f_{1'} +
\sum_{i=2,3;2_0}\frac{V_{1i}f_{i2_0} c_{2_0}}{\E -E_i}.
\label{40}
\er

The results (\ref{37}) and (\ref{38}) are in essence those obtained previously
by  Arima and collaborators \cite{Tak1,Tak2} and by the J\"{u}lich group
\cite{Dro1,Dro2}.
On the other hand, when  terms containing the matrix elements
$\bra{2_0}\hat{d}_{11'}\ket{ 2_0'}$ are neglected in eqs. (\ref{39}) and
(\ref{40}), one finds the results
derived in our previous works \cite{Mar2}.
\footnote{These terms give rise to  disconnected graphs, which are
nonphysical, as well as to  double connected graphs represented in
fig. 1d and 1e, respectively.
As seen from relations (\ref{43})
and (\ref{47}) below, they do not contribute to the
response function.}

In order to elucidate some of the content of these equations we turn next
to a perturbative expansion of the response function and examine the
leading corrections to the
unperturbed $1p1h$ response
$R^0(E)=\sum_1|f_1|^2/(\E - E_1)$.
To achieve maximum simplicity we first omit the residual interaction
within the $1p1h$ sector and backward contributions, so that to second order
the Bethe-Salpeter equation Eq. (\ref{25}) reads
\br
G_{\sf 11'}(E) \cong G_{\sf 11}^0(E) +G_{\sf 11}^0(E) \Sigma_{\sf 11'}(E)
G_{\sf 1'1'}^0(E),
\label{41}
\er
which substituted in  Eq. (\ref{11}) leads to  the desired
approximation for the response function. Within the ESRPA one gets:
\br
&&R(E)\cong R^0(E)
+\sum_{2,2_0,2_0'} c_{2_0}^* \frac{f_{22_0}^*f_{22_0'}}{\E -E_2}c_{2_0'}
+2\sum_{1,2,2_0}\frac{\Re (f_1^*f_{22_0}c_{2_0})}
{\E -E_1} \frac{V_{12}}{\E -E_{2}}
\nonumber \\
&&+ \sum_{1,1'}\frac{f_1^*}{\E - E_1} \left[\sum_{2_0,2_0'}c_{2_0}^*c_{2_0'}
\bra{2_0} \hat{D}_{11'} \ket{ 2_0'} (E - E_1 - E_{2_0})
+ \sum_2 \frac{V_{12} V_{21'}}{\E -E_2}\right] \frac{f_{1'}}{\E -E_{1'}},
\label{42}
\er
and in the ETRPA:
\br
&&R(E)\cong R^0(E)
+\sum_{i=2,3;2_0,2_0'} c_{2_0}^* \frac{f_{i2_0}^*f_{i2_0'}}{\E -E_i}c_{2_0'}
+2\sum_{i=2,3;1,2_0}\frac{\Re (f_1^*f_{i2_0}c_{2_0})}
{\E -E_1} \frac{V_{1i}}{\E -E_i}
\nonumber \\
&&- \sum_{1,1'}\frac{f_1^*}{\E - E_1} \left[\sum_{2_0,2_0'}c_{2_0}^*c_{2_0'}
\bra{2_0} \hat{d}_{11'} \ket{ 2_0'} (E - E_1 - E_{2_0})
- \sum_{i=2,3} \frac{V_{13} V_{31'}}{\E -E_i}\right] \frac{f_{1'}}{\E -E_{1'}}
\label{43}.
\er

Now the two
expressions (\ref{42}) and (\ref{43}) can be shown to be equivalent.
This  results in fact from explicitly performing the sums over $3p3h$
states in Eq. (\ref{43}). To do that one first rewrites
these sums making use of relations (\ref{33}) and (\ref{36}) as:
\br
\sum_{3,2_0,2_0'} c_{2_0}^* \frac{f_{32_0}^*f_{32_0'}}{\E -E_3}c_{2_0'}
= \sum_{3,2_0,2_0'} c_{2_0}^* f_1^*\frac{\bra{2_0}(\hat{D}_{11'} +
\hat{d}_{11'})\ket{2_0'}}{\E -E_1-E_{2_0}}
f_{1'}c_{2_0'},\label{44}
\er
\br
2\sum_{1,3,2_0}\frac{\Re (f_1^*f_{32_0}c_{2_0})}
{\E -E_1} \frac{V_{13}}{\E -E_{3}}= -
2\sum_{1,2_0,2_0'}c_{2_0}^* f_1^* \frac{E_{2_0}
\bra{2_0}(\hat{D}_{11'} + \hat{d}_{11'})\ket{2_0'}}
{(\E - E_1)(\E-E_1-E_{2_0})}
f_{1'}c_{2_0'}.\label{45}
\er
and
\br
\sum_{1,1',3}\frac{f_1^*}{\E -E_1}
\frac{V_{13} V_{31'}}{\E -E_3}\frac{f_{1'}}{\E -E_{1'}}=
\sum_{1,1',2_0,2_0'}
\frac{f_1^* c_{2_0}^*E_{2_0}\bra{2_0}(\hat{D}_{11'} +
\hat{d}_{11'})\ket{2_0'}E_{2_0'}f_{1'}c_{2_0'}}
{(\E -E_1)(\E -E_1-E_{2_0})(\E -E_{1'})},\label{46}
\er
The result of performing the sum is:
\br
\sum_{1,1',2_0,2_0'}\frac{f_1^*}{(\E -E_1)}c_{2_0}^*c_{2_0'}
\bra{2_0}(\hat{d}_{11'}+\hat{D}_{11'})\ket{2_0'}(E -E_1-E_{2_0})
\frac{f_{1'}}{(\E -E_{1'})},
\label{47}
\er
which substituted in Eq. (\ref{43}) gives the expression (\ref{42}) also
for the ETRPA response.

The cancellation among the $3p3h$ on the energy-shell contributions can be
exhibited also making use of the  Rayleight-Schr\"{o}dinger perturbation
expansion, i.e.,
\be
\ket{\tilde{i}}=\ket{i} + \ket{i}^{(1)} + \cdot \cdot \cdot \hspace{3mm}
\mbox{and}~~~ \tilde {E}_i= E_i + E_i^{(1)} + \cdot \cdot \cdot,~i= ipih,
\label{48}
\ee
where the perturbed wave functions and energies are indicated by the
symbol $\sim$ and the superscript points the order of the correction
introduced by the residual interaction $\hat{V}$ on the unperturbed
quantities $\ket{i}$ and $E_i$.
The amplitude for the $\hat{F}$- excitation from the correlated ground
state to the perturbed $3p3h$ states reads
\be
\bra{\tilde{3}}\hat{F}\ket{\tilde{0}}=
\frac{\bra{\tilde{3}}[\hat{H},\hat{F}]\ket{\tilde{0}}}
{\tilde{E}_3-\tilde{E}_0} =
\frac{\bra{3}[\hat{H},\hat{F}]\ket{0}}{E_3 - E_0}+{\cal O}(\hat{V}^2),
\label{49}
\ee
with
\be
\bra{3}[\hat{H},\hat{F}] \ket{0} =
\sum_1 \bra{3}\hat{V}\ket{1}\bra{1}\hat{F}\ket{0}-
\sum_{2_0}\bra{3}\hat{F}\ket{2_0}\bra{2_0}\hat{V}\ket{0}\equiv 0,
\label{50}
\ee
where the last equivalence is a direct consequence of the relations
(\ref{33}), i.e.,
\footnote{Note that $\hat{H}_0$ does not contribute since
$\bra{3} [\hat{H_0},\hat{F}] \ket{0}=0$.}
\be
\sum_1 V_{31} f_1
=-\sum_{1,2_0} c_{2_0}E_{2_0} \ov{1;2_0}{3}f_1
= \sum_{2_0}f_{3,2_0} V_{2_00}
\label{51}
\ee

Thus we see once more that, up to the second order in $\hat{V}$,
the $3p3h$ final states do
not contribute to the response function and that
$\vert\bra{\tilde{3}}\hat{F}\ket{0}\vert^{2}\cong{\cal O}(\hat{V}^4)$.
The Goldstone diagrams for the fourth order $3p3h$ on the mass-shell
contributions to the response function are shown in  figs. 1e and 1f.

At first glance it might look as if the connected Goldstone diagrams
associated with the terms (\ref{44}), (\ref{45}) and (\ref{46}) of the
ETRPA response (illustrated in Figs. 1a, 1b and 1c, respectively) should
give rise to  on the mass-shell $3p3h$ contributions, through the imaginary
part of the propagator $(\E -E_3)^{-1}$. However, Eq. (\ref{47}) shows that
these
contributions in fact cancel out so that the $3p3h$ sector only affects
the $1p1h$ excitations by coupling them with
the virtual intermediate states $\ket{1;2_0}$. Thus in spite of  including
the $3p3h$ propagator in the Green's function, three nucleon ejection does
not occur in the leading order processes. The
above  mentioned diagrams also explain the physical meaning of the fourth term
in the expression (\ref{42}).
The cancellation of  on shell $3p3h$ contributions results from the
destructive interference between amplitudes involving creation of the $3p3h$
state from a ground state correlation and from $V_{31}$ coupling respectively.
A similar calculation in which the backward part of Eq. (\ref{25}) and/or
the residual interaction within the $1p1h$ space are kept up to the
relevant order leads again to the same result. It is worth stressing that
this does not depend on the form of the two-body force used as residual
interaction or on the size of single particle space.

\newpage

\newpage
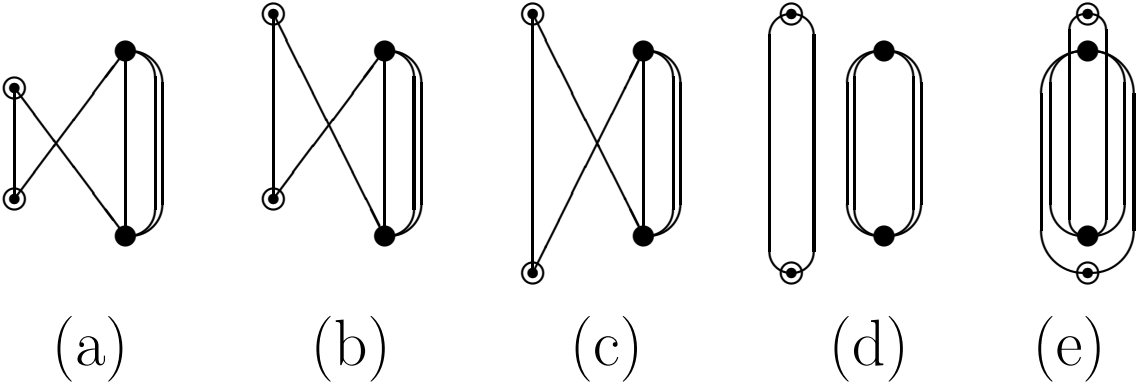
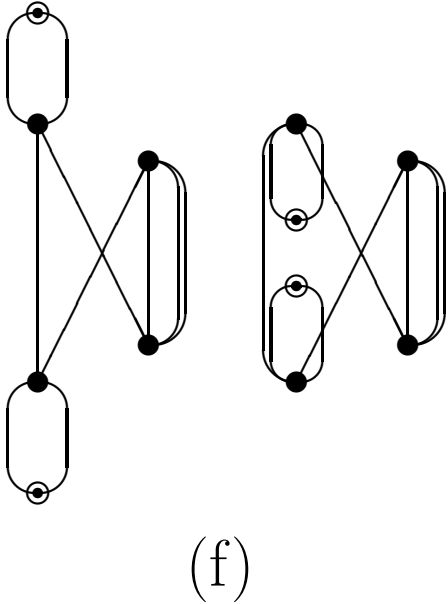
\begin{figure}
\caption{ Graphical representation of the second and fourth order contributions
to the response function. The dotted circles
(\protect $\odot$) denote the one-body vertices and the filled ones
(\protect $\bullet$) indicate the two-body matrix elements.
The diagrams (a), (b) and (c) correspond, respectively, to the terms given by
eqs. (\protect \ref{44}), (\protect \ref{45}) and (\protect \ref{46}).
Second order unlinked and double-linked graphs analogous to the diagram (c)
are shown in figures (d) and (e), respectively.
The last ones, although contained in eqs. (\protect \ref{39}) and
(\protect \ref{40}), do not contribute to the response function.
Finally, figure (f) illustrates the fourth order on the
energy-shell $3p3h$ processes.}
\label{fig}
\end{figure}

\newpage
\pagestyle{empty}
\setlength{\unitlength}{1.4pt}
\begin{picture}(200,300)(70,50)
\thicklines
\multiput (100,300)(70,0){3}{\oval (20,50)[r]}
\multiput (100,300)(70,0){3}{\oval (16,50)[r]}
\multiput (100,325)(70,0){3}{\line (0,-1){50}}
\multiput (100,325)(70,0){3}{\circle*{6}}
\multiput (100,275)(70,0){3}{\circle*{6}}
\multiput (140,335)(70,0){2}{\circle{6}}
\put (210,265){\circle{6}}
\put (70,315) {\circle{6}}
\multiput (70,285)(70,0){2}{\circle{6}}

\multiput (140,335)(70,0){2}{\circle*{3}}
\put (210,265){\circle*{3}}
\put (70,315) {\circle*{3}}
\multiput (70,285)(70,0){2}{\circle*{3}}
\put (210,335){\line (1,-2){30}}
\put (210,265){\line (1, 2){30}}
\put (210,335){\line (0,-1){70}}
\put (70,315){\line (3,-4){30}}
\put (70,285){\line (3,4){30}}
\put (70,315){\line (0,-1){30}}
\put (140,335){\line (1,-2){30}}
\put (140,285){\line (3,4){30}}
\put (140,335){\line (0,-1){50}}
\multiput (305,300)(55,0){2}{\oval (20,50)}
\put (305,300){\oval (16,50)}
\multiput (305,325)(55,0){2}{\circle*{6}}
\multiput (305,275)(55,0){2}{\circle*{6}}
\put (280,335){\circle{6}}
\put (280,265){\circle{6}}
\put (280,335){\circle*{3}}
\put (280,265){\circle*{3}}
\put (280,300){\oval (12,70)}
\put (360,335){\circle{6}}
\put (360,265){\circle{6}}
\put (360,335){\circle*{3}}
\put (360,265){\circle*{3}}
\put (360,305){\oval (10,60)}
\put (360,295){\oval (25,60)}
\put (80,240){{\huge(a)}}
\put (150,240){{\huge(b)}}
\put (220,240){{\huge(c)}}
\put (290,240){{\huge(d)}}
\put (345,240){{\huge(e)}}
\multiput (200,100)(70,0){2}{\oval (20,50)[r]}
\multiput (200,100)(70,0){2}{\oval (16,50)[r]}
\multiput (200,125)(70,0){2}{\line (0,-1){50}}
\multiput (200,125)(70,0){2}{\circle*{6}}
\multiput (200,75)(70,0){2}{\circle*{6}}
\multiput (170,135)(70,0){2}{\circle*{6}}
\multiput (170,65)(70,0){2}{\circle*{6}}
\multiput (170,135)(70,0){2}{\line (1,-2){30}}
\multiput (170,65)(70,0){2}{\line (1, 2){30}}
\put (170,135){\line (0,-1){70}}
\put (240,100){\oval (18,70)[l]}
\put (170,150){\oval (16,30)}
\put (170,50){\oval (16,30)}
\put (170,165){\circle*{3}}
\put (170,165){\circle{6}}
\put (170,35){\circle*{3}}
\put (170,35){\circle{6}}

\put (240,78){\oval (14,26)}
\put (240,122){\oval (14,26)}
\put (240,91){\circle{6}}
\put (240,109){\circle{6}}
\put (240,91){\circle*{3}}
\put (240,109){\circle*{3}}
\put (210,10){{\huge(f)}}
\put (210,-10){Figure 1}
\end{picture}
\end{document}